\documentclass[twocolumn,showpacs,aps,pra]{revtex4}
\usepackage{amsmath}
\usepackage{amsfonts}
\usepackage{amssymb}
\usepackage{graphicx}
\usepackage{hyperref}
\usepackage{bm}
\everymath{\displaystyle} 
\newcommand{\pr}[1]{\ensuremath{\left[#1\right]}} 
\newcommand{\pc}[1]{\ensuremath{\left(#1\right)}} 
\newcommand{\px}[1]{\ensuremath{\left\lbrace#1\right\rbrace}} 
 
\newcommand{\ket}[1]{\ensuremath{\left\vert#1\right\rangle}}

\begin{document}

\title{Controlled ripple texturing of suspended graphene membrane due to
coupling with ultracold atoms}
\author{Sofia Ribeiro}
\affiliation{Quantum Optics and Laser Science, Blackett Laboratory,
Imperial College London, Prince Consort Road, London SW7 2BW, United Kingdom}
\email{s.de-carvalho-ribeiro11@imperial.ac.uk}
\author{Stefan Scheel}
\affiliation{Institut f\"ur Physik, Universit\"at Rostock,
Universit\"atsplatz
3, D-18051 Rostock, Germany}

\date{\today}

\begin{abstract}
We explore the possibility to create hybrid quantum systems that combine
ultracold atoms with graphene membranes. We investigate a setup in which a
cold atom cloud is placed close to a free--standing sheet of graphene at
distances of a few hundred nanometers. The atoms then couple strongly to the
graphene membrane via Casimir--Polder forces.
Temporal changes in the atomic state of the atomic cloud changes the
Casimir--Polder interaction, thereby leading to the creation of a backaction
force in the graphene sheet. This setup provides a controllable way
to engineer ripples in a graphene sheet with cold atoms. 
\end{abstract}


\maketitle


With the advances in trapping and coherently manipulating clouds of ultracold
atoms near microstructured solid-state surfaces \cite{atomchipbook},
the possibility of constructing hybrid atom/solid-state quantum systems have
attracted considerable attention (see,
e.g., Refs.~\cite{Hammerer,PST137_2009_014001,CRP12_871_2011,arXiv1210_4151}). 
Such a hybrid system would consist of ultracold atoms that can be manipulated by
laser light, and a solid-state system that could for instance be controlled by
electrical currents. The influence of the solid-state substrate on the atomic
dynamics is well established; dispersion potentials \cite{acta2008} and their
consequences such as quantum reflection \cite{quantumreflection} and line shifts
\cite{kubler2010} are sufficiently well understood.
However, a backaction of the atom cloud on the solid-state system is rather
challenging. If atoms are regarded as mechanical oscillators
\cite{CRP12_871_2011}, the impedance mismatch due to the large mass
difference between a single atom and a mechanical oscillator limits the
atom--surface coupling. The routes that have been taken so far to alleviate this
discrepancy are either to select a subsystem within the solid-state device
\cite{CRP12_871_2011}, to decrease the effective size of the macroscopic system
\cite{PRA75_062905_2007,Schneeweiss12}, or to enhance the coupling of ions to a
membrane via resonant modes of an optical cavity
\cite{Hammerer,PST137_2009_014001}.
It has been proposed that laser light could be used to couple also the motion
of ultracold trapped atoms to the vibrational modes of a mechanical oscillator. 
In recent experiments, using magnetic \cite{PRL97_227602_2006} or
surface--force coupling \cite{PRL104_143002_2010}, atoms are used to study
vibrations of micromechanical oscillators. In Ref.~\cite{PRL107_223001_2011},
the backaction of the atoms onto the oscillator vibrations as well as the effect
of the membrane vibrations onto the atoms were observed. 

Our goal is to find a coupling mechanism between ultracold atoms and a
solid-state system that is strong enough to provide a mutual interaction between
them. In this article we show that the Casimir--Polder force
\cite{acta2008}, a dispersion interaction that is due to quantum fluctuations of
the electromagnetic field, can provide precisely that. The Casimir-Polder force
on an atom in thermal equilibrium with its environment is typically attractive;
however, in out-of-equilibrium situations such as for atoms prepared in energy
eigenstates the sign of that force can be reversed \cite{thermalCP} and can
reach extremely large values for highly excited (Rydberg) atoms
\cite{AlexPRA2010}. Any cycling transition between the ground state and an
excited state thus translates into an oscillating dispersion force.
The coupling could be increased by minimizing the impedance mismatch using
oscillators with low mass such as carbon nanotubes or graphene membranes. Freely
suspended graphene crystals can exist without a substrate, suspended graphene
flakes or scaffolds have been observed for single layers and bilayers
\cite{NL446_60_2007}. Suspended graphene membranes can be created with diameters
that are comparable to the diameter of a Bose--Einstein condensate.
Free-standing graphene membranes have a key advantage over bulk systems studied
in previous works as the membrane can be cleaned from adsorbates by passing a
current through it \cite{NJP13_083020_2011}.

In the resulting hybrid quantum system, driving the atomic cloud to excited
states could be used to engineer ripples on a graphene membrane. Ripples are an
intrinsic feature of graphene sheets which influence their electronic
properties; perturbations to nearest neighbour hopping might cause the same
effects as inducing an effective magnetic fields and changing local potentials
\cite{PRL97_016801_2006, PRL103_046801_2009, PRL105_156603_2010}. The ability to
control ripple structures could allow a device design based on local
strain and selective bandgap engineering \cite{RPP74_082501_2011}. 
The possibility of constructing an all-graphene circuit, one of the big
goals in graphene science, could be achieved by applying the patterning of
different devices and leads by means of appropriate cuts in the sheet.
In Ref.~\cite{PRL103_046801_2009} it has been proposed to deposit graphene onto
substrates with regions that can be controllably strained on demand, or by
exploring substrates with thermal expansion heterogeneity; the generation of
strain in the graphene lattice is then capable of changing the in-plane hopping
amplitude in an anisotropic way. Controlled ripple texturing using both
spontaneously and thermally generated strains was first reported in
Ref.~\cite{NNL4_562_2009}, where the possibility was shown to control ripple
orientation, wavelength and amplitude by controlling boundary conditions and
making use of graphene's negative thermal expansion coefficient.

In the following, we evaluate the Casimir-Polder force between a single
graphene sheet and a rubidium atom in various energy eigenstates and determine
the minimal number of atoms needed to excite a ripple. 
For the calculation of the interaction potential we assume the
sheet to be infinitely extended, thereby neglecting possible effects that may
arise from the finite extent of the flake. For planar structures, the
Casimir-Polder potential of an atom in an energy eigenstate $|n\rangle$ at a
distance $z_A$ away from the macroscopic body with permittivity
$\varepsilon(\omega)$ can be written as \cite{acta2008}
\begin{gather}
U_{\mathrm{CP}} \pc{z_{A}} = \frac{\hbar \mu_{0}}{8 \pi^{2}} \int_{0}^{\infty}
d \xi\, \xi^{2} \alpha_n(i \xi) \int\limits_{0}^{\infty} d k_{\parallel}
\frac{e^{-2 k_{\parallel} \gamma_{0z} z_{A}} }{\gamma_{0z}} \nonumber \\
\times \left[ \mathrm{R}_{\mathrm{TE}} + \mathrm{R}_{\mathrm{TM}} \left( 1-
\frac{2 k_{\parallel}^{2} \gamma_{0z}^{2} c^{2} }{\xi^{2}} \right) \right] 
\nonumber \\
+ \frac{\mu_{0}}{4 \pi} \sum_{k \neq n} \omega_{n k}^{2} \mathbf{d}_{n k}
\otimes \mathbf{d}_{k n} \int_{0}^{\infty} d \kappa_{0z} e^{-2 \kappa_{0z}}
\nonumber \\
 \times \mathrm{Re} \pr{\mathrm{R}_{\mathrm{TE}} +
\mathrm{R}_{\mathrm{TM}} \pc{1 + \frac{2 \kappa_{0z}^{2}
c^{2}}{\omega^{2}}}  }.  
\label{eq:Ucp}
\end{gather}
where $\gamma_{iz} = \sqrt{1+\varepsilon(i\xi)\xi^{2}/(c^{2} k_\|^2)}$,
$\kappa_{0z} = \sqrt{ k_\|^2+\omega^2/c^2}$ and $\alpha_n(\omega)$ is the atomic
polarizability defined by
\begin{equation}
\mathbf{\alpha}_n(\omega) = \lim_{\varepsilon \rightarrow 0} \frac{2}{\hbar}
\sum_{k \neq n} \frac{\omega_{kn} \mathbf{d}_{nk} \otimes
\mathbf{d}_{kn}}{\omega_{kn}^2-\omega^2-i\omega\varepsilon}\,.
\label{eq:atomicpol}
\end{equation}
The first term in Eq.~(\ref{eq:Ucp}) describes the nonresonant part of the
Casimir-Polder potential, recognisable by the integration along the imaginary
frequency axis, $\omega=i\xi$, whereas the second term is related to resonant 
photon exchange between the atom and the graphene sheet. Equation~\eqref{eq:Ucp} 
is strictly valid only at zero temperature. 
However, we can assume that a potential experiment with ultracold atoms could
be performed at sufficiently low temperatures for thermal excitations to only
play a subordinant role; in addition, the distance of those atoms from the
graphene sheet will be much smaller than the thermal wavelength
$\lambda_T=hc/(k_B T)$. In situations in which either assumption fails to hold, a
replacement of the frequency integral by a Matsubara sum,
\begin{equation}
\frac{\hbar}{\pi}\int\limits_0^\infty d\xi\,f(i\xi)\mapsto
2k_BT\sum\limits_{n=0}^\infty \left(1-\frac{1}{2}\delta_{0n} \right)f(i\xi_n)
\,,
\end{equation}
with Matsubara frequencies $\xi_n=n2\pi k_BT/\hbar$ \cite{thermalCP}, has to be
employed. The material properties of graphene enter via their reflection
coefficients $\mathrm{R}_{\mathrm{TE}}$ and $\mathrm{R}_{\mathrm{TM}}$. Due to
graphene's unique electronic structure, a full calculation of its
electromagnetic reflection coefficients is in fact possible from first
principles. Following Ref.~\cite{PRB84_035446_2011}, in order to derive the
reflection coefficient of a graphene sheet, the dynamics of quasiparticles are
described within the (2+1)--dimensional Dirac model. The quasiparticles in
graphene obey the linear dispersion law $\omega= v_F k$, where $v_F \approx
c/300$ is the Fermi velocity valid for energies below 2~eV
\cite{RMP_Peres_2010}. More elaborate models for the conductivity are not needed
here because at frequencies above the dominant atomic transitions the
polarizability $\alpha_n(i\xi)$ does no longer contribute to the integral in
Eq.~\eqref{eq:Ucp}.

Thermal corrections become important only for $k_B T \gtrsim \Delta$, 
where $\Delta$nis the gap parameter of quasiparticle excitations
\cite{PRA86_012515_2012}. At finite temperature, the potential is well
approximated by inserting the temperature-dependent reflection coefficients in
the lowest term in the Matsubara sum ($j=0$) while keeping the zero-temperature
coefficients for all higher Matsubara terms. However, it has been shown in
Ref.~\cite{PRB82_155459_2010}, that at room temperature the static value of the
polarizability increases only by 10 percent. The TM reflection coefficient
increases by one percent due to finite temperature and the TE coefficient
vanishes altogether at zero frequency. For this reason and with the similar
approach in Ref.~\cite{PRB85_195427_2012} in mind, we have used the
zero-temperature limit of the polarizability as a very good approximation.

The interaction of the quasiparticles with external electromagnetic fields can
be described within this Dirac model. From the boundary conditions of the
electric and magnetic fields one finds the reflection and transmission
coefficients given the specific values of mass gap $m$ and chemical potential
$\mu$. For simplicity we will set $m =\mu=0$ (perfect Dirac cone) for which the
difference between this approximation for suspended graphene samples ($m, \mu
\sim 0.01$eV) is less than $1\%$ \cite{PRB84_035446_2011}. 
One then arrives at the reflection coefficients of a free standing graphene
sheet in vacuum as 
\begin{eqnarray}
\mathrm{R}_{\mathrm{TM}} &=& \frac{4 \pi \alpha \sqrt{k_{0}^{2} +
k_{\parallel}^{2}} }{4 \pi \alpha \sqrt{k_{0}^{2} + k_{\parallel}^{2}} + 8
\sqrt{k_{0}^{2} + \tilde{v}^{2} k_{\parallel}^{2}}}\,, \\
 \mathrm{R}_{\mathrm{TE}} &=& - \frac{4 \pi \alpha\sqrt{k_{0}^{2} +
\tilde{v}^{2} k_{\parallel}^{2}} }{4 \pi \alpha \sqrt{k_{0}^{2} + \tilde{v}^{2}
k_{\parallel}^{2}} + 8 \sqrt{k_{0}^{2} + k_{\parallel}^{2}}} \,,
\label{eq:reflcoefG}
\end{eqnarray}
where we defined $k_{0}^{2} = \xi^{2}/c^{2}$ and $\tilde{v}=v_F/c=(300)^{-1}$;
$\alpha=1/137$ is the fine structure constant.

For a ground-state rubidium atom the force $\mathbf{F}_{\mathrm{CP}}
\pc{z_A} = -\bm{\nabla}_A U_{\mathrm{CP}} \pc{z_A} $ at $z_A=200$~nm is rather
small --- $F_{\mathrm{CP}} (5 S_{1/2}) = -1.05 \times 10^{-22}$~N.
The atom--surface coupling can be vastly enhanced by promoting atoms to highly
excited Rydberg states, i.e. states with very high principal quantum number $n$.

The primary motivation for the study of Rydberg atoms is to take advantage of
the unique opportunities afforded by their exaggerated properties which make
them extremely sensitive to small-scale perturbations of their environment and
to dispersion forces. A second important aspect of atoms in Rydberg states is
their regularity \cite{Gallagher}. For example, the free-space radiative
lifetime of high-lying Rydberg states increases as $n^{3}$. The influence of
macroscopic bodies modifies those atomic relaxation rates \cite{acta2008}.
Earlier results \cite{AlexPRA2010} showed a strong enhancement in the transition
rates of Rydberg atoms near a surface. Intimately connected (via a Hilbert
transform or Kramers--Kronig relation) to the lifetime is the dispersive energy
shift that a Rydberg atom experiences in the vicinity of a macroscopic body
(Casimir-Polder shift) \cite{AlexPRA2010} or in close proximity of another atom
(van der Waals shift). An important consequence of the latter is the Rydberg
blockade effect that prevents multiple Rydberg excitations within a volume of
radius $a_{\mathrm{block}}$, the blockade radius \cite{NPL5_115_2009}.

\begin{figure}[ht]
\includegraphics[width=8cm]{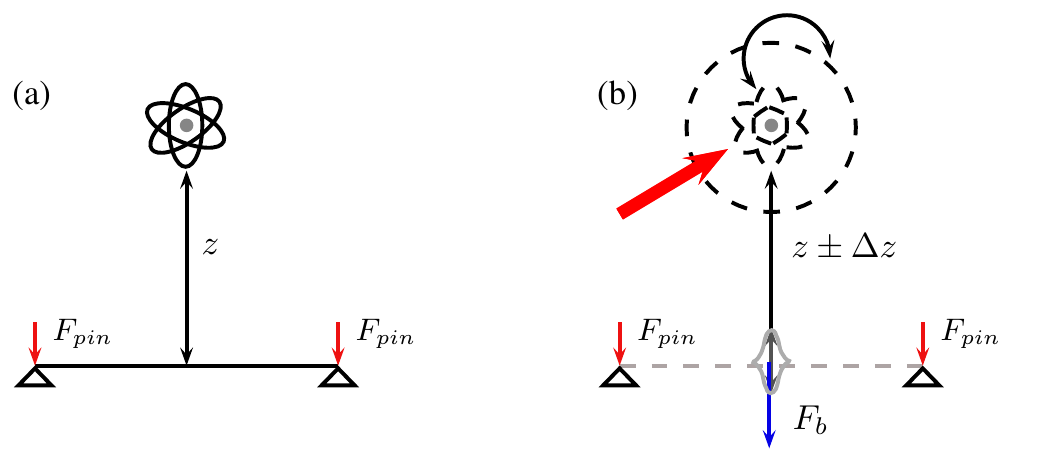}
\caption{(Color online) Schematic diagram (not to scale) showing an atom next
to a suspended graphene membrane. The arrows indicate the two forces $F_{pin}$,
$F_{b}$ at interplay in the system.}
\label{graphene_atom_coupling}
\end{figure}

\begin{table}
\caption{Casimir-Polder force between rubidium atoms in different atomic states
and graphene sheet and the number of atoms needed to excite a ripple of $1$~nm
amplitude. 
\label{table:results}}
\begin{center}
\begin{tabular}{l c c c c}
  \hline
  \hline
  Atomic State & \multicolumn{2}{c}{$T = 0$K} & \multicolumn{2}{c}{$T = 300$K}\\
    & $F_{\mathrm{CP}} (N)$ & $N_{\mathrm{atoms}}$ & $F_{\mathrm{CP}} (N)$ &
$N_{\mathrm{atoms}}$ \\ 
  \multicolumn{5}{c}{$z_{A} = 200$~nm}\\
  \hline
  $\ket{26 S_{1/2}}$ & $2.29 \times 10^{-16}$ & 70 & $-1.89 \times 10^{-15}$ & 9\\
  $\ket{29 S_{1/2}}$ & $3.72 \times 10^{-16}$ & 43 & $-4.08 \times 10^{-15}$ & 4\\
  $\ket{32 S_{1/2}}$ & $5.72 \times 10^{-16}$ & 28 & $-8.15 \times 10^{-15}$ & 2\\
  $\ket{34 S_{1/2}}$ & $7.47 \times 10^{-16}$ & 22 & $-1.25 \times 10^{-14}$ & 2\\
  \hline
  \hline
   \multicolumn{5}{c}{$z_{A} = z_{\mathrm{min}} (n)$}\\
  \hline
  $\ket{26 S_{1/2}}$ & $8.88 \times 10^{-15}$ & 2 & $-7.36 \times 10^{-14}$ & 1\\
  $\ket{29 S_{1/2}}$ & $6.04 \times 10^{-15}$ & 3 & $-6.65 \times 10^{-14}$ & 1\\
  $\ket{32 S_{1/2}}$ & $4.25 \times 10^{-15}$ & 4 & $-6.05 \times 10^{-14}$ & 1\\
  $\ket{34 S_{1/2}}$ & $3.41 \times 10^{-15}$ & 5 & $-5.69 \times 10^{-14}$ & 1\\
  \hline
  \hline
\end{tabular}
\end{center}
\end{table}
In Table~\ref{table:results} we show numerical values of the Casimir-Polder
force acting on rubidium atoms. One observes that the force is attractive for
ground-state atoms, but repulsive for highly excited atoms. This is due to the
increased contributions of the resonant Casimir-Polder force associated with
real-photon transitions as opposed to the nonresonant force components due to
virtual-photon exchanges. A handy feature of Rydberg atoms is thus the
tunability of their interaction strength by choosing a particular Rydberg state
\cite{Gallagher,JPB45_113001_2012}. The excitation into Rydberg states with
principal quantum numbers ranging from $n=20$ up to the ionization threshold are
typically accomplished by a two--photon excitation scheme (for an experimental
review see \cite{JPB45_113001_2012}). Positioning the atom cloud at a fixed
distance away from the surface, one can then excite atoms to the desired Rydberg
state. The backaction of the atoms, mediated by the Casimir-Polder force, onto
the membrane will be a periodic bending force $F_{b}$. Thus, when driving an
atom from its ground state to a Rydberg state and back, one cycles between an
attractive (when the atom is in the ground state) and a repulsive interaction
(when the atom is in the Rydberg state) between atom and graphene sheet. In
addition it is well known that a free--floating graphene sheet would always
crumple at room temperature, hence the need to support the graphene sheet by a
substrate. At very low temperatures, the graphene membrane experiences a
combination of the following forces: (a) the substrate--pinning force $F_{pin}$
that prevents the graphene membrane from sliding and (b) the bending force 
$F_{b}$ due to the Casimir-Polder potential, see
Fig.~\ref{graphene_atom_coupling}.
Measurements on layered graphene sheets of thickness between 2 and 8~nm have
provided spring constants that scale as expected with the dimensions of the
suspended section, and range from 1 to 5~N/m \cite{JVSTB25_6_2007}. Other
experiments studied the fundamental resonant frequencies from electromechanical
resonators made from graphene sheets \cite{Science_315_490_2007}. For mechanical
resonators under tension $T$ the fundamental resonance mode $f_{0}$ is given by 
\begin{equation}
f_{0} = \px{\pr{A \sqrt{\frac{E}{\rho}} \frac{t}{L^{2}}}^{2} + A^{2} 0.57
\frac{T}{\rho L^{2} w t}}^{1/2}
\end{equation}
where $E$ is Young's modulus, $\rho$ is the mass density; $t,\, w, \, L$ are
the thickness, width and length of the suspended graphene sheet and $A$ is a
clamping coefficient ($A$ is equal to 1.03 for doubly clamped beams and 0.162
for cantilevers). The effective spring constant of the fundamental resonance
mode is given  by $\kappa_{\mathrm{eff}}=m_{\mathrm{eff}} \omega^{2}$, where
$m_{\mathrm{eff}}= 0.735 L w t \rho$ \cite{Science_315_490_2007}. 
In the limit of vanishing tension, the fundamental resonance mode is 
$f_{0} = A \sqrt{E/\rho} \: t/L^{2}$. However, we have to assume a finite value
for the tension, for which we choose $T=0.1$~nN. Tension between graphene
and trenches is a random process depending on the production technique and the
interaction with the substrate and for that reason very difficult to control
\cite{Science_315_490_2007}. Using the known values for bulk graphite $\rho=
2200$~kg/m$^{3}$ and $E= 1.0$~TPa, for a graphene cantilever with $t=$0.3~nm,
$L=$3~$\mu$m and $w=$2~$\mu$m the force needed to create a curvature on graphene
with 1~nm amplitude is approximately $16$~fN.  
In order to create a force necessary to produce a ripple of a determined
amplitude --- AFM imaging measures amplitudes in graphene sheets from 0.7 to
30~nm \cite{NNL4_562_2009} --- one has to excite a particular number $N$ of
atoms from the cloud. 

Upon inspection of Table~\ref{table:results} one observes that, for a
cloud of cold $^{87}$Rb atoms at a fixed distance of 200~nm from the graphene
membrane, one would need to excite one or more atoms in order to create a ripple
with 1~nm amplitude. With the blockade radius $a_\mathrm{block}$ of several
micrometers between neighbouring Rydberg atoms only a limited number of atoms
can be resonantly excited to Rydberg states at the same time. If larger numbers 
of Rydberg excitations are needed, the laser line width has to be chosen large enough
to bridge the van der Waals shift between two neighbouring Rydberg atoms. This is 
easily achieved by pulsed-laser excitation with typical line widths of several
$10-100$~MHz.

The interplay between atom--surface distance and principal quantum number $n$ is
of crucial importance in this process. For fixed atom--surface distance, the
interaction increases with $n$ so that fewer atoms are needed to induce a
desired ripple amplitude. However, due to its increasing
size, there is a limit to how close a Rydberg atom can possibly be brought to a
surface, or to what Rydberg state an atom at a given distance can be excited.
This limiting distance can be estimated simply from the classical atomic radius
as $z_\mathrm{min}(n)\gtrsim \sqrt{5}n^2 a_B$ where $a_B$ is the Bohr radius and
the numerical factor $\sqrt{5}$ has been chosen to ensure that its wavefunction
does not overlap with the surface. We see that the number of atoms decreases when 
placed at their minimal distance and, at finite temperature, this number may
decrease to only one atom needed to create a 1~nm ripple.
\begin{table}
\caption{\label{table:number} Minimal number of atoms required to generate a
ripple with 1~nm amplitude. The atoms are assumed to be held at their
respective minimal distances at zero temperature.}
\begin{center}
\begin{tabular}{l c c}
  \hline
  \hline
  Atomic State & $z_\mathrm{min}(n)$ & $N_\mathrm{min}(n)$ \\
  \hline
  \hline
  $\ket{23 S_{1/2}}$ & $62$~nm & 1 \\
  $\ket{30 S_{1/2}}$ & $106$~nm & 3 \\
  $\ket{36 S_{1/2}}$ & $153$~nm & 6 \\
  $\ket{43 S_{1/2}}$ & $218$~nm & 12 \\
  \hline
  \hline
\end{tabular}
\end{center}
\end{table}

An estimate of the number of atoms in a given Rydberg $n$ state needed to generate
a ripple with amplitude 1~nm can thus be obtained as follows. From
Ref.~\cite{AlexPRA2010} we know that the Casimir-Polder force in the
non-retarded limit scales as $F_\mathrm{CP} \propto n^4/z^4$. We then equate the
necessary number of atoms to generate a force of, say $16$~fN, by using the
scaling law on a particular reference state, say $32S$ from
Table~\ref{table:results} at zero temperature. Together with the constraint on
distance, $z_\mathrm{min}(n)\gtrsim n^2(0.118\,\mbox{nm})$, this yields a lower
bound on the number of required atoms as $N_\mathrm{min}(n)\gtrsim 3.6 \times
10^{-6}n^4$. This result seems counterintuitive at first in the sense that
excitation to higher Rydberg states does not seem to increase the force and
lower the number of required atoms. This is due to the competition of
increasing force at fixed distance and larger minimal separation with increasing
$n$. Numerical values for an estimate of the number of atoms needed to be held
at their respective minimal distances are provided in Table~\ref{table:number}.

Realization of the proposed setup requires placing and controlling an atom very
close to a surface. Achieving such control is challenging because atom-surface
forces are comparable with typical trapping forces for cold atoms in this
regime. Atomic ensembles have been stably trapped using magnetic traps formed by
patterned electrodes at distances of 500~nm from a surface
\cite{PRL92_050404_2004, PRL104_143002_2010} and down to 215~nm by using optical
dipole traps based on evanescent waves \cite{PRL104_203603_2010,
PRL109_033603_2012}. In Ref.~\cite{Science340_1202_2013}, a tightly focused
optical tweezer is used to achieve a minimum trap distance of about 100~nm for
realistic laser intensities.

In conclusion, we have shown that it is possible to construct hybrid quantum
systems consisting of cold atoms and solid devices in which a very small number
of atoms exert influence on a much larger object. Here we investigated the
creation of ripples in a graphene membrane due to laser-controlled atom--surface
interactions. Because atoms in different quantum states, in particular highly
excited Rydberg states show vastly different interaction strengths, the
modification of the Casimir-Polder potential creates an effective force on the
graphene sheet. This ability to control and manipulate ripples opens up a number
of novel research possibilities such as the investigation of the effects of
ripples on graphene's electrical and optical properties. 

The key idea in quantum emulators setups with cold gases (bosons, fermions or
mixtures) is to control and simulate other systems of interest, based on the
universality of quantum mechanics. Atom-light interaction can be used to
generate artificial gauge potentials acting on neutral atoms
\cite{RMP83_1523_2011}. In the same way, by tailoring ripples in graphene 
via Casimir--Polder forces is introducing the same effects onto graphene as
those induced by an effective magnetic fields, similarly creating an artificial
gauge potential. This technique also provides a route towards coherent
manipulation of atom-graphene systems. For example, an atom in a coherent
superposition of ground and (highly) excited states leaves the sheet
in a similar superposition of curvatures, thus providing an effective backaction
between cold atoms and a solid-state system that leaves the hybrid system
potentially in an entangled state. We expect such quantum effects only to be
achievable for amplitudes smaller than $1$~nm which have been shown to exist
\cite{Gibertini10}. The advantage of smaller ripples is also the lower number
of atoms for their excitation. Another major advantage of using such a hybrid
system is the fact that we could do a true non-destructive quantum
measurement of the atomic state by testing only the graphene sheet.

S.R. would like to  acknowledge fruitful discussions with V.~Pereira,
F.~Hipolito and R.~Lopes. S.R. is supported by the PhD grant
SFRH/BD/62377/2009 from FCT.

\bibliography{phdthesis}

\end{document}